\documentstyle[11pt]{article}
\font\cero=cmbx10 scaled 1728
\font\uno=cmcsc10 scaled 1200
\font\dos=cmti10 scaled 1200

\title{{\cero From vacuum field equations on principal
bundles to Einstein's equations with fluids}}
\author{{\uno M. Montesinos-Vel\'{a}squez\thanks{Electronic
address :  merced@fis.cinvestav.mx} and T.
Matos\thanks{Electronic address :
tmatos@fis.cinvestav.mx}}\\
{\dos Departamento de F\'{\i}sica} \\
{\dos Centro de Investigaci\'{o}n y de Estudios Avanzados
del IPN} \\  
{\dos Apartado Postal 14-740, 07000,
M\'{e}xico, D.F., M\'exico}}
\date{ }
\begin{document}
\maketitle  
\section*{ }
{\uno Abstract}. In the present work we show that the
Einstein equations on $M$ without cosmological constant and
with perfect fluid as source, can be obtained from the
field equations for vacuum with cosmological constant on
the principal fibre bundle $P\left(\frac1I M,U(1)\right)$,
$M$ being the space-time and $I$ the radius of the internal
space $U(1)$.\\[1ex]
{\uno Resumen}. Mostramos que las ecuaciones de Einstein
sobre $M$ sin constante cosmol\'ogica y con fluido perfecto
como fuente, pueden obtenerse a partir de las ecuaciones de
campo para vac\'{\i}o con constante cosmol\'ogica sobre el
haz fibrado principal $P\left(\frac1I M,U(1)\right)$, donde
$M$ es el espacio-tiempo e $I$ el radio del espacio interno
$U(1)$.\\[2ex]
PACS: 11.10.Kk ; 98.80.Dr

\section*{\uno 1. Introduction}
In a recent work  [1] it has been shown that vacuum
solutions in scalar-tensor theories are equivalent to
solutions of general relativity with imperfect fluid as
source. The above models have the defect that the scalar
fields do not arise from a natural framework of
unification, but they are put by hand as in the
inflationary models [2] and are therefore artificial fields
in the theory. On the other hand, we know that the
geometric formalism of principal fibre bundles [3,4] is a
natural scheme to unify the general relativity theory with
gauge field theories (Abelian and Non-Abelian). If the
principal fibre bundle $P(\tilde M,U(1))$ is endowed with a
metric ``dimensionally reducible" to $\tilde M$ by means of
the reduction theorem [5], {\it i.e.}, if the metric can be
built out from quantities defined only on $M$, then the
scalar fields arise in a natural way. Therefore, it is
important to study the above model in the context of [1]
for the particular principal fibre bundle $P(\frac1I
M,U(1))$, $M$ being the space-time and $I$ the scalar
field. This paper is organized as follows: in the next
section we review the geometric formalism of principal
fibre bundles while in Sect. 3 we deduce the Einstein
equations without cosmological constant and perfect fluid
as source from the field equations on $P(\frac1I M, U(1))$
for vacuum and cosmological constant. We give an example in
Sect. 4 when $\tilde M$ is conformally $FRW$. Finally we
summarize the results in Sect. 5.

\section*{\uno 2. The geometry}
The actual version of the Kaluza-Klein theories is based on
the mathematical stucture of principal fibre bundles [5,6].
In this scheme, the unification of the general relativity
theory with the gauge theories is a natural fact. Moreover,
the reduction theorem provides a metric on (right)
principal fibre bundles $ P(\tilde M, G)$ which is
right-invariant under the action of the structure group $G$
on the whole space $P$. In the trivialization of the bundle
this metric reads [5,6]
\begin{eqnarray}
\mathaccent "705E g = {\tilde
g}_{\alpha\beta}\,dx^{\alpha}\otimes dx^{\beta} +
\xi_{mn}\,(\omega ^m + A^m_{\alpha}\,dx^{\alpha})\otimes
(\omega ^n + A^n_{\beta}\,dx^{\beta}),
\end{eqnarray}
where the metric of the base space $\tilde M$ (generally
identified with the space-time of general relativity) is
${\tilde g}_{\alpha\beta}\,dx^{\alpha}\otimes dx^{\beta}$
while the metric on the fibre $(x^{\alpha}=\mbox{const.})$
is $\xi_{mn}\,\omega ^m \otimes \omega ^n$ and $\{\omega
^m\}$ is a basis of right-invariant 1-forms on $G$. The
quantities ${\tilde g}_{\alpha\beta}$, $\xi_{mn}$ and
$A^n_{\alpha}$ depend only on the coordinates on $\tilde M$
and the $A^n_{\alpha}$ correspond to Yang-Mills potentials
in the gauge theory while the $\xi_{mn}$ are the scalar
fields.

In particular, the principal fibre bundle $P(\tilde
M,U(1))$ has the metric
\begin{eqnarray}
\mathaccent "705E g = {\tilde
g}_{\alpha\beta}\,dx^{\alpha}\otimes dx^{\beta} +
I^2\,(d\psi + A_{\alpha}\,dx^{\alpha})\otimes (d\psi +
A_{\beta}\, dx^{\beta}),
\end{eqnarray}
where the scalar field $I$ correspond to the radius of the
internal space $U(1)$ and $\psi$ is the coordinate on
$U(1)$ too. However, the magnitude of the internal radius
$I$ depends on the particular cases; cosmological or
astrophysical models (for details on units and magnitude on
the scalar field $I$ see Refs. [6,7]). For vanishing
electromagnetic potential, $A_{\alpha} = 0$, we obtain the
unification of ${\tilde g}_{\alpha\beta}$ with the scalar
field $I$
\begin{eqnarray}
\mathaccent "705E g = {\tilde
g}_{\alpha\beta}\,dx^{\alpha}\otimes dx^{\beta} +
I^2\,d\psi^2.
\end{eqnarray}

By using Eq. (3) we compute the Ricci tensor
\begin{eqnarray}
{\mathaccent "705E R}_{\alpha\beta} & = & {\tilde
R}_{\alpha\beta}-I^{-1}\,I_{;\alpha\beta},\\
{\mathaccent "705E R}_{\alpha 4} & = & 0,\\
{\mathaccent "705E R}_{44} & = & -I \,\Box{I},
\end{eqnarray}
where greek indices run on 0,1,2,3 and the label ``4"
corresponds to the fifth dimension.

Usually the base space $\tilde M$ of $P(\tilde M, G)$ is
identified as the space-time , in this paper we adopt the
version where the base space ${\tilde M}$ of $P(\tilde M ,
U(1))$ is conformally the space-time $M$ of general
relativity, {\it i.e.}, $\tilde M = \frac1I M$. That is to
say, we start with the metric (compare Ref. [8])
\begin{eqnarray}
\mathaccent "705E g = \frac1I {{
g}_{\alpha\beta}}\,dx^{\alpha}\otimes dx^{\beta} +
I^2\,d\psi ^2,
\end{eqnarray}
where $g_{\alpha\beta}\,dx^{\alpha}\otimes dx^{\beta}$ is
the space-time metric. Then by using Eq. (7) we obtain the
Ricci tensor 
\begin{eqnarray}
{\mathaccent "705E R}_{\alpha\beta} & = & R_{\alpha\beta} +
\frac12\left( I^{-1} \Box{I}- I_{;\lambda}
I^{;\lambda}\right) g_{\alpha\beta} -\frac32 I^{-2}
I_{;\alpha} I_{;\beta},\\
{\mathaccent "705E R}_{\alpha 4} & = & 0,\\
{\mathaccent "705E R}_{44} & = &
{-{I^2}}\,\Box{I}+I_{;\lambda} I^{;\lambda}.
\end{eqnarray}
In the next we use the signature $(-,+,+,+,)$ for the
space-time metric on $M$.

\section*{\uno 3. Perfect fluid structure}
The field equations on $P(\frac1I M, U(1))$ in vacuum with
cosmological constant $\Lambda$ are given by ${\mathaccent
"705E R}_{AB}-\frac{\mathaccent "705E R}{2}\,{\mathaccent
"705E g}_{AB} = \Lambda \,{\mathaccent "705E g}_{AB}$ or in
equivalent form
\begin{eqnarray}
{\mathaccent "705E R}_{AB} = -\frac23 \Lambda\,
{\mathaccent "705E g}_{AB},
\end{eqnarray}
where $A$,$B$ run on greek indices ${\alpha}$ and 4.

By using Eqs. (8)-(10) and (11) we obtain
\begin{eqnarray}
R_{\alpha\beta} & = & I^{-2}\left(\frac12 I_{;\lambda}
I^{;\lambda}\,g_{\alpha\beta}+\frac32 I_{;\alpha}
I_{;\beta}\right)-I^{-1}\left(\frac12 \Box{I}+\frac23
\Lambda \right) g_{\alpha\beta},\\
\Box{I} & = & \frac1I I_{;\lambda} I^{;\lambda}+\frac23
\Lambda .
\end{eqnarray}

By substituting the field equation for $I$ [Eq. (13)] into
the Ricci tensor  [Eq. (12))] we obtain the next equivalent
equations system
\begin{eqnarray}
R_{\alpha\beta} & = & \frac32 I^{-2}\,I_{;\alpha}\,
I_{;\beta} - I^{-1} \Lambda \,g_{\alpha\beta},\\
\Box{I} & = & \frac1I I_{;\lambda} I^{;\lambda} + \frac23
\Lambda .
\end{eqnarray}

On the other hand, by using the Einstein equations without
cosmological constant,
\begin{eqnarray}
R_{\alpha\beta} & = & T_{\alpha\beta}-\frac{T}{2}
g_{\alpha\beta}
\end{eqnarray}
and Eq. (14), we can define the energy-momentum tensor
associated with the scalar field $I$
\begin{eqnarray}
T_{\alpha\beta} = \frac32 I^{-2}\, I_{;\alpha}\, I_{;\beta}
+\left(-\frac34 I^{-2} I_{;\lambda} I^{;\lambda}+I^{-1}
\Lambda\right) g_{\alpha\beta}.
\end{eqnarray}

This energy-momentum tensor is covariantly conserved,
$T^{\alpha\beta}_{\quad ;\beta} = 0$ as follows from the
field equation for $I$. Finally, by comparing the above
energy-momentum tensor associated with the scalar field $I$
with that of an imperfect fluid
\begin{eqnarray}
T_{\alpha\beta} = \rho U_{\alpha} U_{\beta} + 2 q_{(\alpha}
U_{\beta )} + p\,h_{\alpha\beta}+{\pi}_{\alpha\beta},
\end{eqnarray}
where $\rho$ is the energy density of fluid, $U_{\alpha}$
the velocity, $q_{\alpha}$ the heat flux vector, $p$ the
pressure, $\pi_{\alpha\beta}$ the anisotropic stress tensor
and 
\begin{eqnarray}
h_{\alpha\beta} = g_{\alpha\beta} + U_{\alpha} U_{\beta},
\end{eqnarray}
is the projection orthogonal to the velocity, we conclude
[1]
\begin{eqnarray}
q_{\alpha} & = & 0,\\
\pi_{\alpha\beta} & = & 0,\\
\rho & = & -\frac34 I^{-2} I_{;\lambda}
I^{;\lambda}-\frac{\Lambda}{I},\\
p & = & -\frac34 I^{-2} I_{;\lambda}
I^{;\lambda}+\frac{\Lambda}{I},
\end{eqnarray}
where the velocity has been choosen in the form [1]
\begin{eqnarray}
U_{\alpha} \equiv \ \frac{I_{;\alpha}}{\sqrt{-I_{;\lambda}
I^{;\lambda}}}
\end{eqnarray}

That is to say, Eqs. (20)-(23) implies that Eq. (17) has
the structure corresponding to a perfect fluid. Moreover,
if $\Lambda =0 $ then Eqs. (17) and (20)-(23) correspond to
the so called  ``Zeldovich ultrastiff matter" fluid,
$p=\rho$ (see Ref. [1]).

\section*{\uno 4. Example: the p.f.b. $P(\frac1I FRW,
U(1))$}
We start from the metric 
\begin{eqnarray}
\mathaccent "705E g =
\frac{1}{I(t)}\left[-dt^2+R^2(t)\,\left(\frac{dr^2}
{1-k\,r^2}+r^2\,d\theta^2
+r^2 {\mbox{sin}}^2
\theta\,d\phi^2\right)\right]+I^2(t)\,d\psi^2,
\end{eqnarray}
where $I=I(t)$ on account of the isotropy and homogeneity
of the $FRW$ metric. In this case the Eqs. (14)-(15) read
\begin{eqnarray}
-3 \left(\frac{\mathaccent "7F {R}}{R}\right) & = & \frac32
{\left(\frac{\mathaccent 95 {I}}{I}\right)}^2
+\left(\frac1I\right)\Lambda ,\\
2\left(\frac{k}{R^2}\right)+2{\left(\frac{\mathaccent 95
{R}}{R}\right)}^2+\left(\frac{\mathaccent "7F
{R}}{R}\right) & = & -\left(\frac1I\right)\Lambda ,\\
3\left(\frac{\mathaccent 95
{R}}{R}\right)\left(\frac{\mathaccent 95 {I}}{I}\right) -
{\left(\frac{\mathaccent 95
{I}}{I}\right)}^2+\left(\frac{\mathaccent "7F
{I}}{I}\right) & = & -\frac23\left(\frac1I\right)\Lambda ,
\end{eqnarray}
where dot means derivation with respect to the cosmological
time $t$. These equations are equivalent to the Einstein
equations for $FRW$ with perfect fluid as source, provided
that
\begin{eqnarray}
\rho & = & \frac34{\left(\frac{\mathaccent 95
{I}}{I}\right)}^2-\left(\frac1I\right)\Lambda ,\\
p & = & \frac34{\left(\frac{\mathaccent 95
{I}}{I}\right)}^2+\left(\frac1I\right)\Lambda .
\end{eqnarray}

By the way, the field equation for $I$ [Eq. (28)] is the
covariant conservation of $T_{\alpha\beta}$,
$T^{\alpha\beta}_{\quad ;\beta}=0$ 
\begin{eqnarray}
\mathaccent 95 {\rho} + 3\left(\frac{\mathaccent 95
{R}}{R}\right)\left(\rho +p\right) = 0.
\end{eqnarray}

\section*{\uno 5. Conclusion}
We have shown that the field equations with cosmological
constant $\Lambda$ on the principal fibre bundle $P(\frac1I
M, U(1))$ are equivalent to the Einstein equations without
cosmological constant on $M$ and with perfect fluid as
source. In order to show it we start from the field
equations on $P(\frac1I M,U(1))$, ${\mathaccent "705E
R}_{AB}=-\frac23\Lambda \,{\mathaccent "705E g}_{AB}$ and
separate them in their 4-dimensional and five dimensional
parts. We have found that from the 4-dimensional part of
these equations it is possible to define an effective
energy-momentum tensor $T_{\alpha\beta}$ and that it is
covariantly conserved, being $T^{\alpha\beta}_{\quad
;\beta} = 0$ equivalent to the field equation for $I$.
Finally, we applied the above result to the particular
bundle $P(\frac1I FRW, U(1))$.
 
\section*{\uno Acknowledgments}
The authors are grateful to CONACyT for financial support.

\section*{\uno References}
\newcounter{ref}
\begin{list}{\hspace{1.3ex}\arabic{ref}.\hfill}
{\usecounter{ref} \setlength{\leftmargin}{2em}
\setlength{\itemsep}{-.98ex}}
\item L.O. Pimentel, {\it Class. Quantum Grav.}\ {\bf 6}
(1989) L263.
\item E.W. Kolb and M.S. Turner, {\it The Early Universe},
Addison-Wesley, Redwood City, CA (1990).
\item A. Trautman, {\it Rep. Math. Phys.}\ {\bf 1} (1970)
29.
\item Y. Choquet-Bruhat, C. De Witt-Morette and the M.
Dillard-Bleick, {\it Analysis, Manifolds and Physics}, Part
I, Revised Edition, North-Holland (1991).
\item C. Choquet-Bruhat and C. De Witt-Morette, {\it
Analysis, Manifolds and Physics}, Part II, North-Holland
(1989), Sect.\ V.13.
\item T. Matos and J.A. Nieto, {\it Rev. Mex. F\'{\i}s.}\
{\bf 39} No. Sup. 1 (1993) 81.
\item D. Bailin and A. Love, {\it Rep. Prog. Phys.}\ {\bf
50} (1987) 1087.
\item P.S. Wesson and J. Ponce de Le/'on, {\it J. Math.
Phys.}\ {\bf 33} (11) (1992) 3883. 
\end{list}
\end{document}